\documentclass[runningheads]{llncs}
\usepackage{graphicx}
\usepackage{tikz}
\usepackage{comment}
\usepackage{amsmath,amssymb} 
\usepackage{color}
\usepackage{hyperref}

\usepackage{array}
\newcolumntype{P}[1]{>{\centering\arraybackslash}p{#1}}

\newcommand{\R}{\mathbb{R}}

\newcommand{\myparagraph}[1]{\smallskip\noindent\textbf{#1}}

\usepackage{float}

\usepackage{sidecap}

\usepackage{booktabs}
\usepackage{multirow}
\newcommand{\ra}[1]{\renewcommand{\arraystretch}{#1}}

\newsavebox\CBox
\def\textBF#1{\sbox\CBox{#1}\resizebox{\wd\CBox}{\ht\CBox}{\textbf{#1}}}

\begin{document}
\def\IPMISubNumber{240}

\title{TopoTxR: A Topological Biomarker for Predicting Treatment Response in Breast Cancer}

\setlength{\textfloatsep}{8pt}
\setlength{\floatsep}{8pt}
\setlength{\belowcaptionskip}{-3pt}

\titlerunning{TopoTxR: A Topological Biomarker for pCR prediction}
%
\author{Fan Wang\orcidID{0000-0002-5402-5065}\and
Saarthak Kapse\and
Steven Liu\and\newline 
Prateek Prasanna\orcidID{0000-0002-3068-3573}\and
Chao Chen\orcidID{0000-0003-1703-6483}}
\authorrunning{Fan Wang et al.}
%
\institute{Stony Brook University, Stony Brook, NY 11794, USA \\
\email{fanwang1@cs.stonybrook.edu},~
\email{saarthak.kapse@stonybrook.edu}\\
\email{\{steven.h.liu, prateek.prasanna, chao.chen.1\}@stonybrook.edu}}
\maketitle              

\begin{abstract}
\label{sec:abstract}
Characterization of breast parenchyma on dynamic contrast-enhanced magnetic resonance imaging (DCE-MRI) is a challenging task owing to the complexity of underlying tissue structures. Current quantitative approaches, including radiomics and deep learning models, do not explicitly capture the complex and subtle parenchymal structures, such as fibroglandular tissue. In this paper, we propose a novel method to direct a neural network's attention to a dedicated set of voxels surrounding biologically relevant tissue structures. By extracting multi-dimensional topological structures with high saliency, we build a topology-derived biomarker, \emph{TopoTxR}. We demonstrate the efficacy of \emph{TopoTxR} in predicting response to neoadjuvant chemotherapy in breast cancer. Our qualitative and quantitative results suggest differential topological behavior of breast tissue on treatment-na\"ive imaging, in patients who respond favorably to therapy versus those who do not.
\footnote{This work was partially supported by grants NSF IIS-1909038, CCF-1855760, and NCI 1R01CA253368-01.
This work used the Extreme Science and Engineering Discovery Environment (XSEDE) \cite{towns2014xsede} Bridges-2 at the Pittsburgh Supercomputing Center through allocation TG-CIS210012, which is supported by NSF ACI-1548562.}


\end{abstract}

\section{Introduction}
\label{sec:intro}
Traditional cancer imaging biomarker studies have mostly been focused on texture and shape-based analysis of the lesion, often ignoring valuable information harbored in the tumor microenvironment. There is an overwhelming evidence of diagnostic and prognostic information in the tumor periphery, such as the peritumoral stroma and parenchyma~\cite{braman2017intratumoral}.
In breast cancer, the phenotypic heterogenity in the extra-tumoral regions stems from factors such as stromal immune infiltration, vascularity, and a combination of fatty and scattered fibroglandular tissue. 
Breast density, composition of fibroglandular tissue, and background parenchymal enhancement have been shown to be associated with breast cancer risk and are also implicated in differential response to therapy~\cite{king2011background}. 
There is hence a need for novel interpretable quantitative approaches to comprehensively characterize breast cancer biology by interrogating the tumor microenvironment and the surrounding parenchyma as observed on routine imaging scans.

Radiomic approaches have been recently used to learn diagnostic and prognostic signatures from breast tumor and surrounding peritumoral regions. Although promising, radiomic features cannot explicitly model the complex and subtle parenchymal tissue structures. Therefore, the learning outcome lacks sufficient interpretability; one cannot derive actionable knowledge regarding the tissue structures from the learnt diagnostic/prognostic models.
\begin{figure}[btp]
\begin{center}
\includegraphics[width=.9\textwidth]{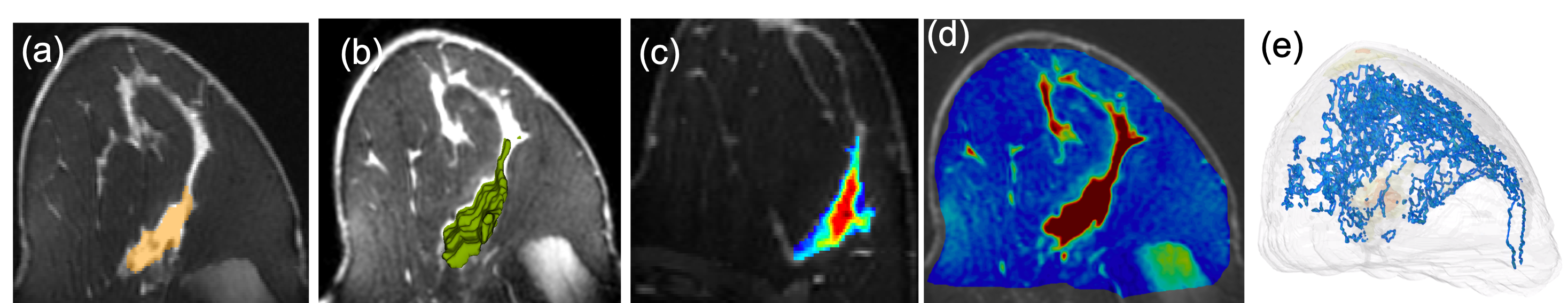}
\caption{ An example MRI image (a) and different radiomics features such as (b) tumor 3D shape, (c) intratumoral texture (Haralick entropy), and (d) whole breast texture (Haralick energy). In (e), we show topological structures from \textit{TopoTxR}, capturing the density of fibroglandular tissue.}
\label{fig:features}
\end{center}
\end{figure}
Convolutional neural networks (CNNs), on the the other hand, have shown great promise in various domains, as they learn feature representations in an end-to-end manner. For breast cancer, CNN models trained on mammography images have shown very strong diagnostic efficacy~\cite{abdelhafiz2019deep}. However, mammograms are of relatively low resolution and are only 2D projections of 3D tissue structures. The loss of true topology and geometry of the 3D tissue structures fundamentally limits the power of mammography-based models. CNN models have been proposed for MRIs, which can characterize the true 3D tissue structures~\cite{selvikvaag2018overview}. Such models are capable of learning features that combine lower level abstractions and high order details which maximally discriminate between different classes. While promising, these methods take whole breast MRI as direct input; a large portion of the input volume may be biologically irrelevant and even noisy enough to significantly bias the prediction task. Besides, 3D CNNs have millions of parameters, and require a large amount of training data which is often unavailable for controlled clinical trials such as the I-SPY1 trial (less than 250 cases)~\cite{newitt2016multi}. CNNs also suffer from the limitation of feature interpretability as they lack direct grounding with the breast tissue structures.

\footnote{This work is accepted for presentation at IPMI 2021.}
 
We present a novel topological biomarker for breast DCE-MRI. Our method bridges the two extremes (hand-crafted imaging features vs.~completely data-driven CNNs). The key idea is to \textbf{direct the model's attention to a much smaller set of voxels surrounding tissue structures with high biological relevance}. This way, the deep convolutional network can be efficiently trained with limited MRI data. Meanwhile, the learning outcome has the potential of connecting to the biological cause manifested on the tissue structure topology. As shown in Figure~\ref{fig:features}, our topological descriptor (e) directly models the breast parenchymal tissue structures, whereas other features (b-d) do not.

To explicitly extract tissue structure is a challenging task. Training a segmentation model may not be always feasible due to the lack of ground truth. Instead, we propose  an unsupervised approach to extract the tissue structures using the theory of persistent homology~\cite{edelsbrunner2010computational}. 
Our method extracts 1D and 2D topological structures (loops and bubbles) with high saliency. These structures correspond to curvelinear tissue structures (e.g., ducts, vessels, etc.) and voids enclosed by tissues and glands in their proximity. We consider these topological structures a reasonable approximation of the tissue structures and combine them with the original MRI image as the input to train 3D CNNs.  
By focusing on such tissue structures and their periphery, we can effectively train a 3D CNN even with small datasets. Additionally, the tissue-centric representation can be effectively visualized for better interpretation.

Although our approach is domain-agnostic, as a use case we focus on  predicting treatment response (TxR) in breast cancer treated with neoadjuvant chemotherapy (NAC). Correct prediction of pathological complete response (pCR) prior to NAC administration can help avoid ineffective treatment that introduces unnecessary suffering and costs. However, reliably predicting pCR using treatment-naive DCE-MRI still remains a challenge with current clinical metrics and techniques. Our method, called \emph{TopoTxR}, significantly outperforms existing methods, including radiomics and image-only CNNs, on the I-SPY1 dataset \cite{newitt2016multi}.

Persistent homology has been used in various biomedical image analysis tasks \cite{hu2019topology,hu2021topology,clough2020topological,dey2019road,wu2017optimal,wang2020topogan}. However, most existing approaches focus on only using the persistence diagrams as direct features. Meanwhile, the topological structures uncovered through the algorithm carry rich geometric information that has not been explicitly utilized. \textbf{Our approach leverages topological structures based on the persistent homology to explicitly direct the attention of convolutional neural networks.} The topology-driven attention enables the CNNs to learn efficiently. Our method outperforms various baselines, including ones that use persistence diagram features. A Python implementation of TopoTxR can be found at our GitHub repository: \url{https://github.com/TopoXLab/TopoTxR}.

\subsection{Related Work}
\label{sec:relatedwork}



Quantitative imaging features have been used in conjunction with machine learning classifiers for prediction of pCR \cite{radiomic_app1,radiomic_app3}. Radiomics approaches, involving analysis of quantitative attributes of tumor texture and shape, have shown promise in assessment of treatment response.  In particular, such features capture appearance of the tumors and, more recently, peritumor regions~\cite{radiomic1,peritumoral1}. Such approaches are often limited by their predefined nature, lack of generalizability, dependency on accurate lesion segmentation, and inability to explain phenotypic differences beyond the peritumoral margin. 
CNNs have been previously applied to breast DCE-MRI for pCR prediction  \cite{CNN1,CNN4,CNN5}.  Owing to the sub-optimal performance of image-only models, image based CNN approaches have been fused with non-imaging clinical variables in order to bolster prediction~\cite{fusheng}.

\myparagraph{Topological information,} in particular, persistent homology, has been used in various image analysis tasks, such as cardiac image analysis \cite{wu2017optimal}, brain network analysis \cite{lee2012persistent}, and neuron image segmentation \cite{hu2019topology}. In recent years, it has been combined with deep neural networks to enforce topological constraints in image segmentation tasks \cite{hu2019topology,clough2020topological}.
Abundant work has been done to learn from information represented by persistence diagrams, e.g., via vectorization \cite{adams2017persistence}, kernel machines \cite{reininghaus2015stable,kusano2016persistence,carriere2017sliced}, or deep neural networks \cite{hofer2017deep}.
However, the topological structures associated with the diagrams, e.g., cycles and bubbles, have not been explored. These structures describe geometric details of the breast tissue (e.g. fibroglandular tissue) and can be mapped to the original breast volume to provide an explicit attention mechanism for CNNs.


\section{Methodology}
\label{sec:method}
\begin{figure}[btp]
\begin{center}
\includegraphics[width=.98\textwidth]{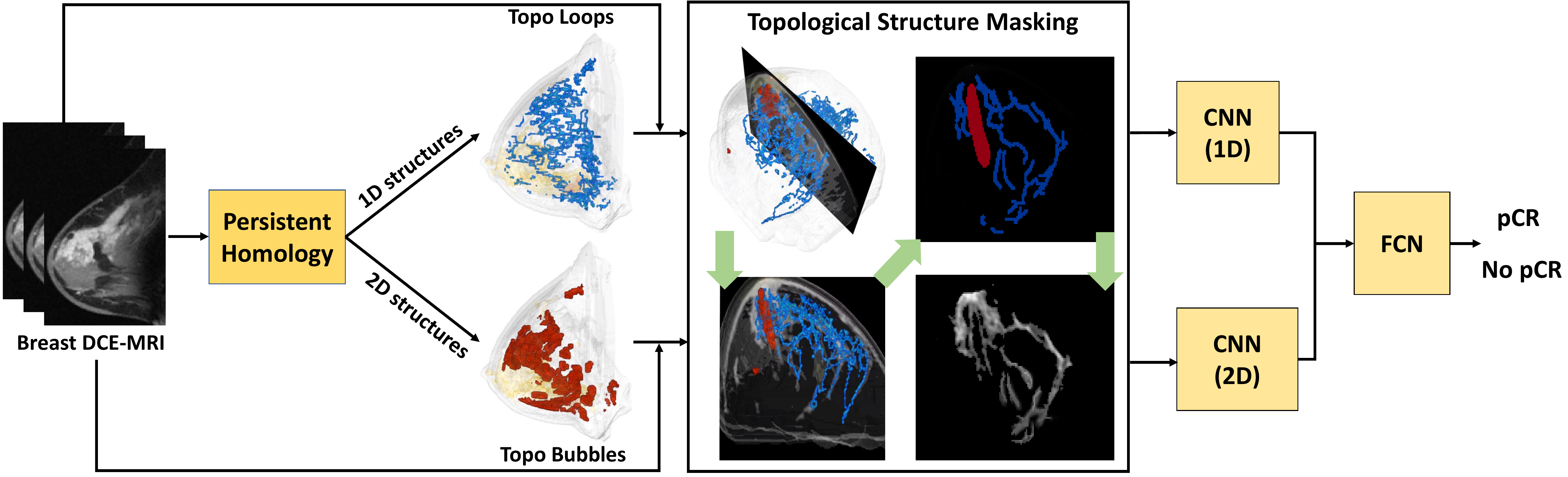}
\caption[overview]{Our proposed TopoTxR pipeline. We extract 1D and 2D topological structures from breast MRI with persistent homology. We create new images in which voxels belonging to these structures have their intensity value from the original breast MRI, and zero otherwise. We create two images corresponding to 1D and 2D topological structures. We use two 3D CNNs and a fully connected network for pCR prediction.}
\label{fig:overview}
\end{center}
\end{figure}

We propose a topological approach to (1) extract topological structures with high saliency as an approximation of the tissue structures, and (2) use the extracted topological structures as explicit attention to train a deep convolutional network. 
Our method is summarized in Fig.~\ref{fig:overview}. 

We first compute salient topological structures from the input image based on the theory of persistent homology \cite{edelsbrunner2010computational}.
Topological structures of dimensions 1 and 2, i.e., loops and bubbles, can both correspond to important tissue structures. 1D topological structures capture curvelinear structures such as ducts, vessels, etc. 2D topological structures represent voids enclosed by the tissue structures and their attached glands.
These topological structures directly delineate the critical tissue structure with high biological relevance. Thus we hypothesize that by focusing on these tissue structures and their affinity, we will have relevant contextual information for prediction.

Next, we propose a topological-cycle-driven CNN to learn from MRIs and the discovered topological structures. We explicitly mask the input MRI image so that only voxels of the extracted topological structures and their vicinity regions are visible. We then train a 3D CNN on this masked image. By focusing on the tissue structure vicinity region, we can train the CNN effectively even with a limited training set. 
We note that there are two types of relevant topological structures, loops and bubbles. Our network consists of two separate 3D CNNs, treating the two types of topological structures separately. The feature representation from the two convolutional networks are concatenated and are provided to fully connected layers for the prediction (pCR vs. no pCR). As will be shown empirically, both types of topology capture complementary structural signatures and are necessary to achieve the best predictive performance. 

In this section, we will first explain the background knowledge about persistent homology. Next, we introduce cycles representing the topological structures. Finally, we will describe our topological-cycle-driven CNN.

\subsection{Background: Persistent Homology}
We review basics of persistent homology in this section. Interested readers may refer to \cite{edelsbrunner2010computational} for more details.
Persistent homology extracts the topological information of data observed via a scalar function. Given an image domain, $X$, and a real-valued function $f:X\rightarrow \R$, we can construct a sublevel set $X_{t} =
\{x \in X: f(x) \leq t\}$ where $t$ is a threshold controlling the ``progress"
of sublevel sets. The family of sublevel sets $\mathcal{X} = \{X_{t}\}_{t \in \mathbb{R}}$ defines a filtration, i.e., a family of subsets of $X$ nested with respect to the inclusion: $X_{\alpha} \subseteq X_{\beta}$ if $\alpha \leq \beta$. As the threshold $t$ increases from $-\infty$ to $+\infty$, topological structures such as connected components, handles, and voids appear and disappear. The birth time of a topological structure is the threshold $t$ at which the structure appears in the filtration. Similarly, the death time is the threshold at which the structure disappears. Persistent homology tracks the topological changes of sublevel sets $X_{t}$ and encodes them in a \emph{persistence diagram}, i.e., a point set in which each point $(b,d)$ represents a topological structure born at $b$ and killed at $d$.

\begin{figure}[btph]
\begin{center}
\includegraphics[width=.96\textwidth]{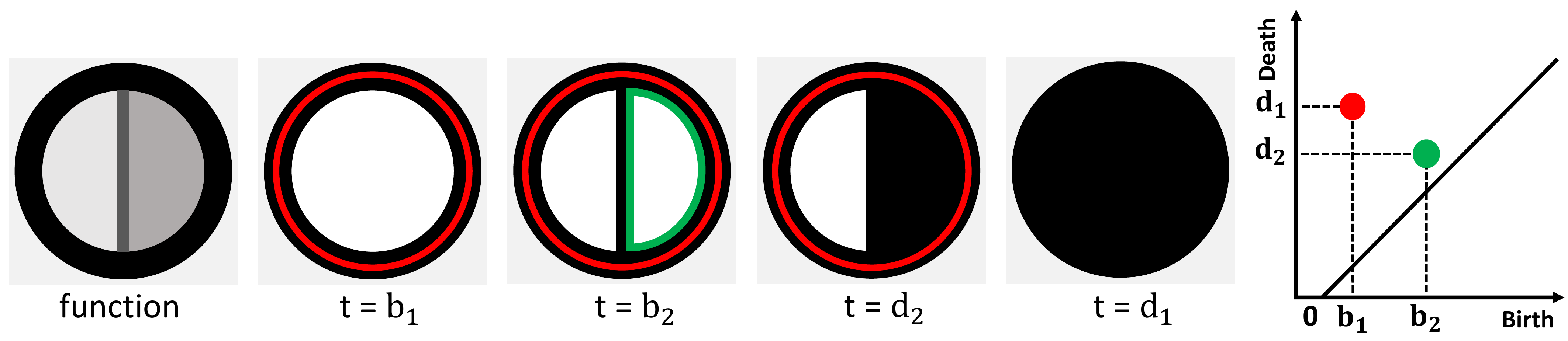}
\caption[ph]{From left to right: a synthetic image $f$, sublevel sets at thresholds $b_{1} < b_{2} < d_{2} < d_{1}$ (in black), and the persistence diagram of dimension 1. The red loop represents a 1D structure born at $b_1$ and killed at $d_1$. The green loop represents a 1D structure born at $b_{2}$ and killed at $d_2$. They correspond to the red and green dots in the diagram. }
\label{fig:ph}
\end{center}
\end{figure}

See Fig.~\ref{fig:ph} for an example function $f$ and its sublevel sets at different thresholds. At time $b_1$, a new handle (delineated by the red cycle $c_{1}$) is created. This handle is later destroyed at time $d_{1}$. Another handle delineated by the green cycle $c_{2}$ is created and killed at $b_{2}$ and $d_{2}$ respectively. The topological changes are summarized in a persistence diagram on the right. Each handle corresponds to a 2D dot in $\mathbb{R}^{2}$, whose $x$ and $y$ coordinates are birth and death times. The difference between a dot's birth and death times is called its \emph{persistence}. 


\subsection{Persistence Cycles and Their Computation}
In this section, we introduce cycles that represent topological structures discovered by persistent  homology. We also present an algorithm to compute these cycles.
Intuitively, a topological cycle of dimension $p$ is a $p$-manifold without boundary. A 1-dimensional (1D) cycle is a loop (or a union of a set of loops). A 2-dimensional (2D) cycle is a bubble (or a union of a set of bubbles). 
A cycle $z$ represents a persistent homology structure if it delineates the structure at its birth. For example, in Fig.~\ref{fig:ph}, the red and the green loops represent the handles born at time $b_1$ and $b_2$, respectively.

We assume a discretization of the image domain into distinct elements, i.e., vertices (corresponding to voxels), edges connecting adjacent vertices, squares, and cubes.
These elements are 0-, 1-, 2-, and 3-dimensional cells.
Any set of $p$-cells is called a \emph{$p$-chain}.
We define the \emph{boundary operator} of a $p$-cell, $\sigma$, as the set of its $(p-1)$ faces.
The boundary of an edge is its two vertices. The boundary of a square consists of the 4 edges enclosing it. The boundary of a cube consists of the 6 squares enclosing it. 
For any $p$-chain, $c$, its boundary is  
$\partial(c) = \sum_{\sigma\in c} \partial(\sigma)$, under mod-2 sum. 
For a set of edges forming a path, its boundary are the two end vertices. For any set of squares forming a patch, its boundary is the loop enclosing the patch. The boundary of a set of cubes is the bubble enclosing them.

\begin{figure}[btph]
\begin{center}
\includegraphics[width=.88\textwidth]{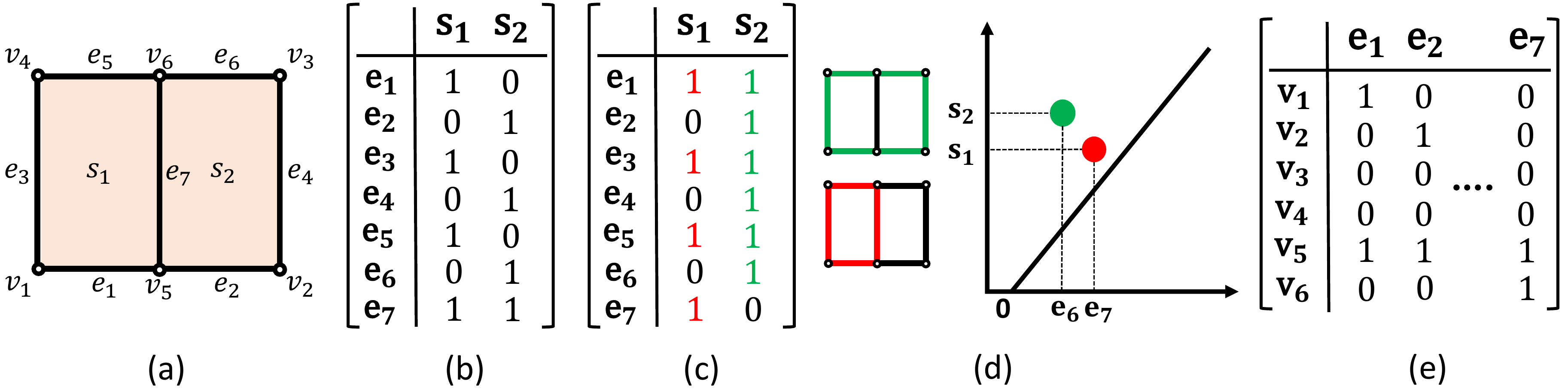}
\caption[cycle]{(a) Example of a cubical complex whose cells are sorted according to the function values. (b) 2D Boundary matrix $\partial$. (c) Reduced boundary matrix. (d) Persistence diagram and resulting cycles corresponding to $\partial$. (e) 1D boundary matrix.}
\label{fig:cycle}
\end{center}
\end{figure}

A $p$-chain is a \emph{$p$-cycle} if its boundary is empty.
All $p$-cycles form the null space of the boundary operator, i.e. $\{c : \partial(c)=\emptyset\}$. Any topological structure, formally defined as a \emph{homology class}, can be delineated by different cycles, which are equivalent to each other in terms of topology. We can choose any of them to represent this class. In persistent homology, for each dot in the persistence diagram, we can represent it with one representative cycle at its birth. In Fig.~\ref{fig:ph}, the red and green cycles represent the two corresponding handles.
Note that the choice of representative cycle is not unique. A relevant question is to choose the shortest representative cycle (i.e., one with the least number of edges) for each dot in the diagram \cite{wu2017optimal,zhang2019heuristic}. In this paper, we focus on choosing a standard representative cycle, leaving the optimal cycle for future work.

\myparagraph{Computation of persistent homology and representative cycles.}
We assume a filtration function on a discretization of the image domain.
An example discretization of a 2D image is given in Fig.~\ref{fig:cycle}(a). We first sort all cells in increasing order according to their function values. The computation of persistence diagrams is then performed by encoding the $p$-dimensional boundary operator in binary matrices named \emph{boundary matrices}, $\partial_p$. $\partial_p$ maps $p$-cells to their boundaries. Fig.~\ref{fig:cycle} shows the 1D and 2D boundary matrices of the given complex and its filtration. The 1D boundary matrix is essentially the incidence matrix of the underlying graph (Fig.~\ref{fig:cycle}(e)). 
High dimensional boundary matrices are defined similarly (e.g., a 2D boundary matrix in Fig.~\ref{fig:cycle}(b)).

Persistence diagram is computed by reducing the boundary matrix similar to Gaussian elimination, but without row or column perturbation. We reduce through column operations performed on $\partial$ from left to right. 
Fig.~\ref{fig:cycle}(c) shows the reduced 2D boundary matrix. 
Once the boundary matrices are reduced. Each non-zero column corresponds to a persistent dot in the diagram. The reduced column itself is the cycle representing the corresponding topological structure. In this paper, we pay attention to both 1D and 2D cycles, corresponding to loops and bubbles. The extracted cycles will be used to explicitly guide 3D CNNs for analysis.
The computation of topological cycles is of the same complexity as the computation of persistent homology. In theory, it takes $O(n^\omega)$ time ($\omega \approx 2.37$ is the exponent in the matrix multiplication time, i.e., time to multiply two $n\times n$ matrices) \cite{milosavljevic2011zigzag}. Here $n$ is the number of voxels in an image. In practice, to compute all cycles of an input image ($256^3$), it takes approximately 5 minutes.
\subsection{Topological-Cycle-Driven 3D CNN}
An overview of our topological-cycle-driven 3D CNN has been provided in the beginning of Section \ref{sec:method}.
Here we describe the technical details.

To compute persistence and topological cycles, we invert the MRI image $f = -I$ so that the tissue structures correspond to low intensity.
After the computation, we select topological cycles representing dots of the diagram with high persistence. 
The general belief is that low-persistence dots tend to be due to noise. Thus we only select high-persistence cycles, which are considered more salient structures, and more likely to represent true tissue structures. The threshold is a hyperparamter tuned in practice. 

Next, we create two binary 3D masks representing 1D and 2D topological cycles, respectively. Both masks are dilated slightly in order to cover both the structures and their vicinity. 
Instead of directly using these binary masks for pCR prediction, we fill the foreground voxels with their original image intensity values. In other words, we mask the input image with the complement of the cycle mask. See\emph{Topological Structure Masking} step in Fig.~\ref{fig:overview} for the masked MRI image. We generate masked images for both 1D and 2D, and provide them to two CNNs.
All masked MRIs are padded to the same size of 256 $\times$ 256 $\times$ 256.

We use separate networks with the same architecture for cycles and bubbles. The CNN consists of 5 3D convolution layers, each followed by a batch normalization layer and a LeakyReLU. The output feature maps from these two 3D CNNs are reshaped and concatenated into a feature vector. This feature vector is sent into a fully connected (FC) network with three FC layers for final pCR prediction. Besides ReLU and batch normalization, a dropout layer is added to the second FC layer. The final output is a vector of size 2. All three networks are trained together in an end-to-end fashion with stochastic gradient descent (SGD) as the optimizer and cross-entropy as loss.
\section{Experimental Results}
\label{sec:experiment}
\begin{table}[t]
\small
\begin{center}
\caption{Comparisons of proposed method against baseline methods on four metrics: accuracy, AUC, specificity, and sensitivity. The p-values in the last row are computed between baseline MRI and TopoTxR.} 
\label{table:mainres}
\ra{1.1}
\begin{tabular}{@{}r | P{2cm} P{2cm} P{2cm} P{2cm} @{}}
\toprule
\multicolumn{1}{r}{} 
& \multicolumn{1}{c}{\textBF{Accuracy}} 
& \multicolumn{1}{c}{\textBF{AUC}} 
& \multicolumn{1}{c}{\textBF{Specificity}} 
& \multicolumn{1}{c}{\textBF{Sensitivity}}\\
\cmidrule{2-5}
\multicolumn{1}{r}{} & \multicolumn{4}{c}{Without Feature Selection}\\
\midrule
Radiomics & 0.517$\pm{0.086}$ & 0.536$\pm{0.098}$ & 0.557$\pm{0.058}$ & 0.477$\pm{0.176}$\\
PD & 0.529$\pm{0.071}$ & 0.537$\pm{0.078}$ & 0.543$\pm{0.075}$ & 0.515$\pm{0.151}$\\
Radiomics+PD & 0.533$\pm{0.080}$ &0.538$\pm{0.095}$ & 0.567$\pm{0.065}$ & 0.5$\pm{0.175}$ \\

\midrule
\multicolumn{1}{r}{} & \multicolumn{4}{c}{With Feature Selection}\\ \midrule
Radiomics & 0.563$\pm{0.085}$ & 0.593$\pm{0.098}$ & 0.552$\pm{0.180}$ & 0.575$\pm{0.081}$\\
PD & 0.549$\pm{0.081}$ & 0.567$\pm{0.097}$ & 0.551$\pm{0.167}$ & 0.547$\pm{0.071}$\\
Radiomics+PD & 0.563$\pm{0.093}$ & 0.587$\pm{0.099}$ & 0.592$\pm{0.178}$ & 0.534$\pm{0.087}$\\

\midrule
\multicolumn{1}{r}{} & \multicolumn{4}{c}{3D CNN}\\ \midrule
MRI & 0.633$\pm{0.200}$ & 0.621$\pm{0.102}$ & 0.570$\pm{0.322}$ & 0.673$\pm{0.354}$\\
TopoTxR (MRI+Topo) & \textbf{0.851}$\pm{0.045}$ & \textbf{0.820}$\pm{0.035}$ & \textbf{0.736}$\pm{0.086}$ & \textbf{0.904}$\pm{0.068}$\\
p-value & 0.0625 & 0.0625 & 0.3750 & 0.1875\\

\bottomrule
\end{tabular}
\end{center}
\end{table}

We validate our method on the task of pCR prediction with ISPY-1 post-contrast DCE-MRI data~\cite{newitt2016multi}. A total of 162 patients are considered - 47 achieving pCR (mean age = 48.8 years), 115 non-pCR (mean age = 48.5 years).
All experiments were performed in a 5-fold cross-validation setting.  
The performance was evaluated with accuracy, area under curve (AUC), specificity, and sensitivity. 
Both mean and standard deviation are reported. 
For all methods, hyperparameters (learning rate, momentum, weight decay factor, batch size, and dropout rate for the dropout layer) are tuned using a grid search, and are selected from a 3-fold cross validation on a small set held out for validation. 



\begin{table}[t]
\small
\begin{center}
\caption{Ablation study results. All numbers are reported from 5-fold cross validations.}
\label{table:abl}
\ra{1.1}
\begin{tabular}{@{}r | P{2cm} P{2cm} P{2cm} P{2cm} @{}}
\toprule
\multicolumn{1}{r}{} 
& \multicolumn{1}{c}{\textBF{Accuracy}} 
& \multicolumn{1}{c}{\textBF{AUC}} 
& \multicolumn{1}{c}{\textBF{Specificity}} 
& \multicolumn{1}{c}{\textBF{Sensitivity}}\\
\cmidrule{2-5}
\multicolumn{1}{r}{} & \multicolumn{4}{c}{Persistence Threshold}\\ \midrule
$90\%$ Remain& 0.826$\pm{0.069}$ & 0.783$\pm{0.063}$ & 0.675$\pm{0.1115}$ & 0.891$\pm{0.084}$\\
$60\%$ Remain& \textbf{0.851}$\pm{0.021}$ & 0.793$\pm{0.028}$ & 0.647$\pm{0.073}$ & \textbf{0.939}$\pm{0.017}$\\

\midrule
\multicolumn{1}{r}{} & \multicolumn{4}{c}{Dimension}\\ \midrule
Dimension 1 & 0.718$\pm{0.068}$ & 0.697$\pm{0.025}$ & 0.639$\pm{0.149}$ & 0.754$\pm{0.161}$\\
Dimension 2 & 0.756$\pm{0.036}$ & 0.691$\pm{0.013}$ & 0.520$\pm{0.116}$ & 0.863$\pm{0.103}$\\

\midrule
\multicolumn{1}{r}{} & \multicolumn{4}{c}{Dilation Radius}\\ \midrule
Radius 2 & 0.721$\pm{0.036}$ & 0.673$\pm{0.024}$ & 0.569$\pm{0.037}$ & 0.777$\pm{0.055}$\\
Radius 4 & 0.677$\pm{0.023}$ & 0.603$\pm{0.007}$ & 0.442$\pm{0.063}$ & 0.764$\pm{0.054}$\\
Radius 8 & 0.646$\pm{0.034}$ & 0.569$\pm{0.040}$ & 0.399$\pm{0.057}$ & 0.737$\pm{0.033}$\\

\bottomrule
\end{tabular}
\end{center}
\end{table}

We compare with various baseline methods.
\textbf{Radiomics}: We compute a 92 dimensional radiomic signature~\cite{pyradiomics} and train a classifier on it. 
Features are extracted solely from the tumor region.
\textbf{PD}: We train a classifier using persistence-diagram-based features, i.e., features extracted from persistence diagrams (PDs) of the input MRI images. While various classifier options are available and behave similarly, we use the 
sliced Wasserstein kernel distance for PDs as a feature vector \cite{carriere2017sliced}. 
\textbf{Radiomics+PD}: We combine both radiomics and PD features and train a classifier on them. 
\textbf{With feature selection}: We apply feature selection to all aforementioned methods, using Mutual Information Difference (MID) and Mutual Information Quotient (MIQ). For all baseline features, we search exhaustively among all combinations of feature selection schemes and a set of classifiers (Random Forests, Linear Discriminant Analysis, Quadratic Discriminant Analysis and SVM). We  report the best results.
\textbf{MRI}: we directly apply a 3D CNN to the original DCE-MRIs.

\myparagraph{Quantitative results.}
Radiomics and PD features yielded better performance when used together with a Random Forest classifier (Table \ref{table:mainres}).
We observe that direct application of a 3D CNN (method MRI) does not perform well, presumably due to the lack of sufficient amount of data.  Our proposed approach (\textit{TopoTxR}: MRI+Topo) outperforms all baseline methods.  Due to the imbalance in the dataset, we also report the classifier specificity and sensitivity.  
Further evaluation, to address data imbalance, will be carried out in future work.


\begin{figure}[btp]
\begin{center}
\includegraphics[width=.81\textwidth]{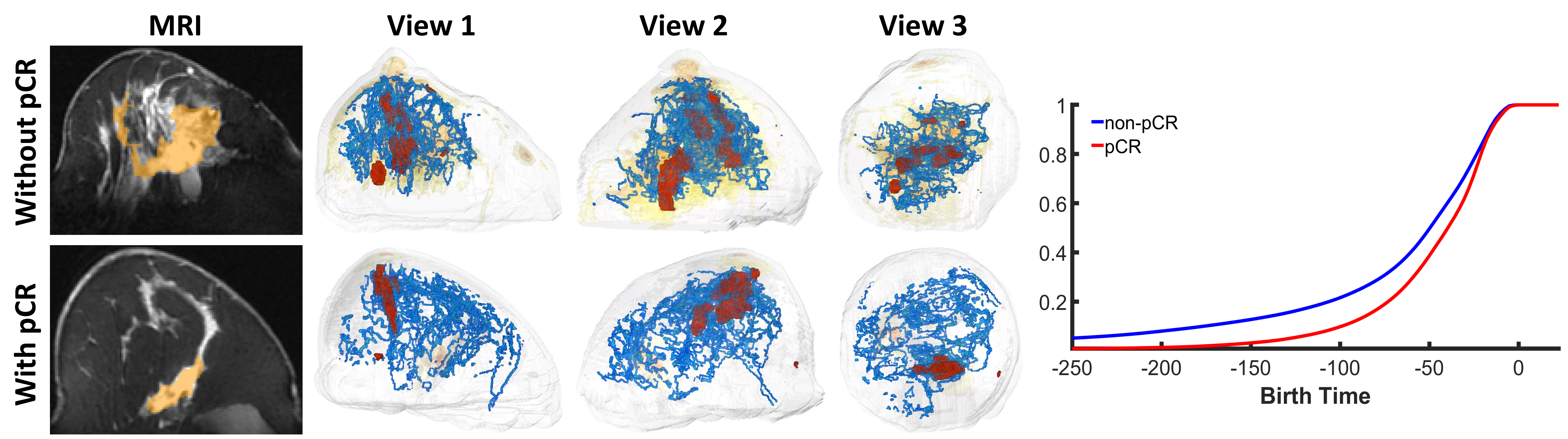}
\caption[vis-struct-new]{Qualitative comparison of patients with and without pCR. First column: Slices of breast DCE-MRI with tumor masked in orange (tumor masks are not used in building TopoTxR). Columns 2-4: 3D renderings of topological structures from three different views. 1-D structures (loops) are rendered in blue and 2-D structures (bubbles) in red. Top row: no pCR, Bottow row: pCR. Right: cumulative density function of topological structures' birth times.}
\label{fig:qualitative}
\end{center}
\end{figure}

\myparagraph{Ablation study.}
Recall that the persistence of a topological structure is defined as the difference between its birth and death times. We threshold out topological structures with low persistence, as they are generally caused by noise and could negatively influence the results. We explore the impact of persistence thresholding by choosing 3 different thresholds so that $90\%$, $60\%$, and $30\%$ of the structures remain. According to Table \ref{table:abl}, retaining $30\%$ structures (refer to TopoTxR's results in Table \ref{table:mainres}) yielded an optimal trade-off between the quantity and quality of the topological structures.
We also tested the method using 1D structures (loops) only and 2D structures (bubbles) only. Both are better than baseline methods, but still inferior to \textit{TopoTxR}. This shows that the 1D and 2D structures provide complementary predictive information.
Finally, the topology structures are dilated to form a mask. We ran an ablation study with regard to the dilation radius and obtain the best performance when no dilation is performed. The results of \textit{TopoTxR} in Table $\ref{table:mainres}$ is reported with $30\%$ structures remaining using combined $1$ and $2$D structures without dilation.
\subsection{Discussion and TopoTxR Feature Interpretation}
The topological structures extracted by TopoTxR capture the breast tissue structures.
Learning directly on these structures and their vicinity provides the opportunity for interpreting the learning outcomes and drawing novel biological insights. 
Here we provide some visual analysis as a proof of concept.

Fig.~\ref{fig:qualitative} shows the TopoTxR topographical structures from different views for a representative DCE-MRI scan from each group. We observe that the structures (1D and 2D) are sparse for the case exhibiting pCR, and are relatively dense for the non-pCR case. In the corresponding MRI images, we note that the pCR breast has scattered fibroglandular breast density with minimal background parenchymal enhancement. The non-pCR breast has a more heterogenous fibroglandular breast density with moderate background parenchymal enhancement. This possibly suggests that the \textit{TopoTxR} features capture the complex fibrogladular structure which can be a potential indicator of treatment response. 

We also compare the topological behavior of the two populations. Recall the birth time of a topological structure is the threshold at which a cycle appears. In our experiments, since we use the inverse image $f=-I$, the birth time essentially captures -1 times the brightness of a structure. In Fig.~\ref{fig:qualitative} (right), we plot the cumulative density function (CDF) of the birth time of topological structures for pCR (red) and non-pCR (blue) patients. The CDFs suggest that pCR patients' tissue structures are generally less bright (or less visible) compared with that of non-pCR patients. This is consistent with our observation on qualitative examples. 
A Kolmogorov-Smirnov test \cite{massey1951kolmogorov} is performed to compare these CDFs. The computed \textit{p-value} is 0.0002, indicating a significant difference between the distributions of birth times of the pCR and non-pCR patient groups.

\section{Conclusion}
This paper presents a novel topological biomarker, \emph{TopoTxR}, that leverages the rich geometric information embedded in structural MRI and enables improvement in downstream CNN processing. In particular, we compute 1D cycles and 2D bubbles from breast DCE-MRIs with the theory of persistent homology; these structures are then used to direct the attention of neural networks. We further demonstrate that \emph{TopoTxR} on treatment-naive imaging is predictive of pCR. 



\bibliographystyle{abbrv}
\bibliography{main}

\end{document}